\documentclass[conference]{IEEEtran}
\usepackage[left=0.67in,right=0.67in,top=0.64in,bottom=0.74in]{geometry}
\usepackage[T1]{fontenc}
\usepackage[latin9]{inputenc}
\usepackage{amsthm}
\usepackage{amsmath} 
\usepackage{graphicx} 
\usepackage{color} 
\usepackage{cite}
\usepackage{algpseudocode}
\usepackage{algorithm}
\usepackage{amssymb}
\usepackage{subfigure}
\usepackage{esint}
\usepackage{algpseudocode}
\usepackage{epstopdf}
\usepackage{multirow}
\usepackage{makecell}
\usepackage{float}
\usepackage{lipsum}
\usepackage{balance}
\usepackage{ulem}

\newtheorem{definition}{Definition}
\newtheorem{remark}{Remark}

\newtheorem{example}{Example}

\theoremstyle{plain}
\theoremstyle{plain}
\newtheorem{theorem}{Theorem}

\newcommand{\comment}[1]{}

\IEEEoverridecommandlockouts

\begin{document}

\title{\vspace{-0.1em} Read-and-Run Constrained Coding for \\ Modern Flash Devices}

\author{
   \IEEEauthorblockN{Ahmed Hareedy\IEEEauthorrefmark{1}, Simeng Zheng\IEEEauthorrefmark{2}, Paul Siegel\IEEEauthorrefmark{2}, and Robert Calderbank\IEEEauthorrefmark{1}}
   \IEEEauthorblockA{\IEEEauthorrefmark{1}Electrical and Computer Engineering Department, Duke University, Durham, NC 27708 USA \\ \IEEEauthorrefmark{2}Electrical and Computer Engineering Department, University of California, San Diego, La Jolla, CA 92093 USA \\ ahmed.hareedy@duke.edu, sizheng@ucsd.edu, psiegel@ucsd.edu, and robert.calderbank@duke.edu}
}
\maketitle

\begin{abstract}

The pivotal storage density win achieved by solid-state devices over magnetic devices in 2015 is a result of multiple innovations in physics, architecture, and signal processing. One of the most important innovations in that regard is enabling the storage of more than one bit per cell in the Flash device, i.e., having more than two charge levels per cell. Constrained coding is used in Flash devices to increase reliability via mitigating inter-cell interference that stems from charge propagation among cells. Recently, capacity-achieving constrained codes were introduced to serve that purpose in modern Flash devices, which have more than two levels per cell. While these codes result in minimal redundancy via exploiting the underlying physics, they result in non-negligible complexity increase and access speed limitation since pages cannot be read separately. In this paper, we suggest new constrained coding schemes that have low-complexity and preserve the desirable high access speed in modern Flash devices. The idea is to eliminate error-prone patterns by coding data only on the left-most page while leaving data on all the remaining pages uncoded. Our coding schemes work for any number of levels per cell, offer systematic encoding and decoding, and are capacity-approaching. Since the proposed schemes enable the separation of pages, we refer to them as \textit{read-and-run (RR)} constrained coding schemes as opposed to schemes adopting \textit{read-and-wait} for other pages. We analyze the new RR coding schemes and discuss their impact on the probability of occurrence of different charge levels. We also demonstrate the performance improvement achieved via RR coding on a practical triple-level cell Flash device.

\end{abstract}

\section{Introduction}\label{sec_intro}

The history of constrained coding dates back to 1948, when Shannon represented a constrained sequence via a finite-state transition diagram (FSTD) and derived the capacity under a constraint \cite{shan_const}. Run-length-limited (RLL) codes were introduced by Tang and Bahl in 1970 to support the evolution of magnetic recording at that time \cite{tang_bahl}, and these codes were based on lexicographic indexing. In 1973, Cover presented a result about enumerative coding \cite{cover_lex} that will prove fundamental for the design of constrained codes based on lexicographic indexing decades later. Among other researchers, Franaszek developed constrained codes based on finite-state machines (FSMs) derived from FSTDs \cite{franaszek}. In 1983, Adler, Coppersmith, and Hassner introduced a systematic method to develop constrained codes based on FSMs \cite{ach_fsm}. Details about the history of constrained coding until 1998 are in \cite{immink_surv}.

Because of their ability to improve performance via eliminating error-prone data patterns and undesirable sequences, constrained codes have a plethora of applications. They find application in one-dimensional (1D) magnetic recording devices, both the old ones, which are based on peak detection, and the modern ones, which are based on sequence detection \cite{vasic_prc, ahh_loco}. They can also be combined with robust signal detection using machine learning \cite{zheng_prnn}. They find application in the emerging two-dimensional (2D) magnetic recording devices as well \cite{wood_tdmr, bd_tdmr}. Moreover, constrained codes are used to achieve DC balance and self-calibration in optical recording devices \cite{immink_opt} in addition to many computer standards for data transmission \cite{saade_comp}.

In Flash devices, charge propagation from cells programmed to high charge levels into cells programmed to lower charge levels is the main reason behind inter-cell interference (ICI) \cite{lee_ici}. This is correct for any number $q$ of charge levels per cell. Mitigating ICI results in remarkable lifetime gains in Flash as demonstrated in \cite{veeresh_mlc} for multi-level cell (MLC) Flash ($q=4$). There are data patterns that are considered usual suspects for contributing most to ICI. Coding to eliminate data patterns resulting in consecutive levels $(q-1)0(q-1)$ was considered in \cite{ravi_const} and \cite{chee_ici}. Coding to eliminate data patterns resulting in consecutive levels or level patterns $(q-1)\mu(q-1)$, for all $\mu < q-1$, was presented in \cite{veeresh_mlc}, \cite{chee_ici}, and \cite{ahh_qaloco}.

A number of recent results revisited \cite{tang_bahl} and \cite{cover_lex} in order to produce efficient constrained codes based on lexicographic indexing, and one example is \cite{braun_lex}. Another example is \cite{ahh_loco}, in which we introduced binary symmetric lexicographically-ordered constrained (S-LOCO) codes and demonstrated density gains in a modern magnetic recording system. We extended our result to single-level cell (SLC) Flash ($q=2$) \cite{ahh_aloco} then to Flash with any number $q$ of levels per cell \cite{ahh_qaloco}. Moreover, we devised a general method to design LOCO codes for any finite set of patterns to forbid \cite{ahh_general}, which will be useful in this paper. We studied the power spectra of binary LOCO codes in \cite{jes_psd}. LOCO codes are capacity-achieving, simple, and easily reconfigurable \cite{ahh_qaloco, ahh_general}.

While the constrained codes in \cite{chee_ici} and \cite{ahh_qaloco} are quite efficient in terms of rate, they require all Flash pages to~be processed together, which negatively affects the access speed. In this paper, we propose \textit{read-and-run (RR)} constrained~coding schemes that allow pages to be accessed separately in modern Flash devices, thus preserving high access speed. There are techniques in the literature that allow page separation; however, they are either incurring notable rate loss \cite{veeresh_mlc} or designed for a specific Flash setup \cite{ravi_const}. Our RR coding schemes incur minor rate loss and work for any Flash device. The key idea is that the constrained code is applied only on one page, while no coding is applied on the other $\log_2 q-1$ pages. We present a 2D RR coding scheme as well as a 1D RR coding scheme that is based on LOCO codes. We study various aspects about these schemes, including the charge-level probabilities. We introduce experimental results in a practical triple-level cell (TLC) Flash device ($q=8$) that demonstrate notable lifetime gains achieved by our coding schemes. 

The rest of the paper is organized as follows. In Section~\ref{sec_map2d}, we discuss the detrimental patterns, the Flash mapping, and our 2D RR coding scheme. In Section~\ref{sec_rrloco2}, we introduce our 1D RR-LOCO coding scheme. In Section~\ref{sec_compropag}, we study the rate, complexity, and error propagation of the new schemes. In Section~\ref{sec_resultstlc}, we present the experimental results on TLC Flash. In Section~\ref{sec_concl}, we conclude the paper.

\section{Patterns, Mapping, and 2D RR Coding}\label{sec_map2d}

As implied in the introduction, literature works do not strictly agree on the set of forbidden patterns to operate on. Additionally, as the Flash device ages, the set of error-prone patterns is expected to expand \cite{ahh_qaloco}. Based on our recent experimental tests on a practical TLC Flash device, we decided to focus on the set characterized as follows. Let
\begin{equation}
\beta_1, \beta_2 \in \mathcal{V}_0 \triangleq \left \{\frac{q}{2}, \frac{q}{2}+1, \dots, q-1 \right \},
\end{equation}
where $q$ is the number of levels per Flash cell (a positive power of $2$) and $\mathcal{V}_1 = \{0, 1, \dots, q-1\} \setminus \mathcal{V}_0$. Then, the set of interest is the set resulting in the level patterns in $\mathcal{L}_q$:\footnote{Levels are defined through their indices $\{0, 1, \dots, q-1\}$ for simplicity.}
\begin{equation}\label{eqn_forb}
\mathcal{L}_q \triangleq \{\beta_1\mu\beta_2, \forall \beta_1,\beta_2 \text{ } \vert \text{ } 0 \leq \mu < \min(\beta_1,\beta_2)\}.
\end{equation}
This set already subsumes all $3$-tuple forbidden patterns adopted in the literature for Flash. A block inside the Flash device can be seen as a 2D grid of wordlines and bitlines, with a cell being placed at each intersection \cite{veeresh_mlc}. Level patterns in $\mathcal{L}_q$ are detrimental whether they occur on $3$ adjacent cells along the same wordline or along the same bitline.

\begin{example}
Consider an MLC Flash device, i.e., $q=4$. In this case, we have $\beta_1, \beta_2 \in \{2,3\}$. Then, the set of interest is the set resulting in:
\begin{equation}
\mathcal{L}_4 = \{202, 212, 203, 213, 302, 312, 303, 313, 323\}.
\end{equation}
The last three elements in $\mathcal{L}_4$ are quite known \cite{veeresh_mlc, ravi_const, ahh_qaloco}.
\end{example}

\begin{algorithm}
\caption{Recursive Alternate Gray Mapping}
\begin{algorithmic}[1]
\State \textbf{Input:} Number of levels per cell $q$, and $p=\log_2 q$.
\State Define $\mathrm{map}$, a binary array of dimensions $q \times p$.
\State Set $\mathrm{map}(0,:) = \bold{1}^p$. \textit{(a sequence of $p$ $1$'s)}
\State \textbf{for} $i \in \{0, 1, \dots, p-1\}$ \textbf{do}
\State \hspace{2ex} \textbf{for} $j \in \{0, 1, \dots, 2^i-1\}$ \textbf{do}
\State \hspace{4ex} $\mathrm{map}(2^i+j,:) = \mathrm{map}(2^i-1-j,:)$.
\State \hspace{4ex} Flip the bit $\mathrm{map}(2^i+j,i)$. \textit{(each sequence in $\mathrm{map}$ is indexed from right to left by $0, 1, \dots, p-1$)}
\State \hspace{2ex} \textbf{end for}
\State \textbf{end for}
\State \textbf{Output:} Array $\mathrm{map}$ that maps each index to binary data.
\end{algorithmic}
\label{alg_ragm}
\end{algorithm}

Next, we discuss how to map from data to charge levels in Flash and vice versa. Since we are interested in page separation throughout this work, the mapping here is from a charge level out of $q$ possible ones to $\log_2 q$ binary bits, one for each page, and vice versa. Gray mapping offers the advantage that there is only one-bit difference between any two adjacent levels, which is valuable for error performance. We adopt a recursive alternate Gray mapping (RAGM), and Algorithm~\ref{alg_ragm} shows how to produce it for any $q$. We highlight that RAGM has already been used in the literature in MLC Flash \cite{veeresh_mlc} and TLC Flash \cite{ravi_const}. Thus, RAGM is not strictly a new contribution.

\begin{example}
Consider a TLC Flash device, i.e., $q=8$. In this case, the output of Algorithm~\ref{alg_ragm}, which is RAGM, becomes:
\begin{align}\label{eqn_map}
0&\longleftrightarrow111, \hspace{+3.0em} 1\longleftrightarrow110, \nonumber \\
2&\longleftrightarrow100, \hspace{+3.0em} 3\longleftrightarrow101, \nonumber \\
4&\longleftrightarrow001, \hspace{+3.0em} 5\longleftrightarrow000, \nonumber \\
6&\longleftrightarrow010, \hspace{+3.0em} 7\longleftrightarrow011.
\end{align}
\end{example}

Now, we are ready to discuss coding schemes. Let us first index the Flash pages the same way the bits in each sequence in the array $\mathrm{map}$ are indexed (see Algorithm~\ref{alg_ragm}). This means that the left-most page is the one indexed by $p-1$. From (\ref{eqn_forb}) and Algorithm~\ref{alg_ragm}, the level patterns in $\mathcal{L}_q$ correspond to binary patterns where the left-most page(s) always has~(have) two $0$'s separated by some bit, i.e., $0x0$. Based on that, forbidding $\{000,010\}$ on the left-most page(s) guarantees that no level pattern in $\mathcal{L}_q$ would appear while writing to a Flash device, with any $q > 2$, at least in one direction. This corresponds to an interleaved RLL $(d,k)=(0,1)$ constraint~\cite{siegel_nasit2015}. Notably, no coding on any other page is needed. Data will therefore be read from each page independently, and immediately passed to the low-density parity-check (LDPC) decoder. This idea is the key idea of our RR constrained coding schemes.{\footnote{An equivalent scheme was proposed for MLC Flash, i.e., $q=4$, in \cite{siegel_nasit2015}.}
 
RR coding can be performed in the wordline direction only (1D), the bitline direction only (1D), or both directions (2D). Observe that such coding will also prevent benign level patterns, e.g., $555$ and $676$ in TLC Flash, resulting in inevitable rate loss. However, as we shall see in Section~\ref{sec_compropag}, this rate loss is small, and RR-LOCO codes are capacity-approaching.

We start here with our scheme for 2D RR constrained coding. As the name suggests, we want to prevent the patterns in $\{000,010\}$ from appearing at the left-most pages in both wordline and bitline directions through simple encoding and decoding. The encoding follows the rules:
\begin{enumerate}
\item On wordlines with indices congruent to $0$ or $1$ (mod $4$), you are allowed to write $0$'s and $1$'s freely in bit positions congruent to $0$ or $1$ (mod $4$) at the left-most pages.
\item On wordlines with indices congruent to $2$ or $3$ (mod $4$), you are allowed to write $0$'s and $1$'s freely in bit positions congruent to $2$ or $3$ (mod $4$) at the left-most pages.
\item In the other bit positions, you can only write $1$'s on wordlines at the left-most pages.
\end{enumerate}

This 2D RR constrained coding scheme is depicted in Fig.~\ref{fig_1}. It is clear from the figure that the patterns in $\{000,010\}$ are eliminated from the left-most pages, which forbids all level patterns in $\mathcal{L}_q$, in both directions. Upon encoding, input data bits are freely placed at the positions marked by $x$ and directly at the other pages. Upon decoding, information at the positions marked by $1$ is omitted, and data bits at the remaining positions are read with no additional processing and with no correlation between different Flash pages.{\footnote{An equivalent 2D scheme forbidding patterns $\{101, 111\}$ on the right-most pages in both worline and bitline directions in MLC Flash was proposed in~\cite{siegel_nasit2015}.}

This 2D scheme is ideal in terms of complexity, access speed, and error propagation (see Section~\ref{sec_compropag}). It might also seem notably better than any 1D scheme in terms of performance. However, as we shall justify later, 1D schemes can achieve almost the same performance with higher rates.

\begin{figure}
\vspace{-0.5em}
\center
\includegraphics[trim={1.35in 2.2in 1.4in 1.0in}, width=3.5in]{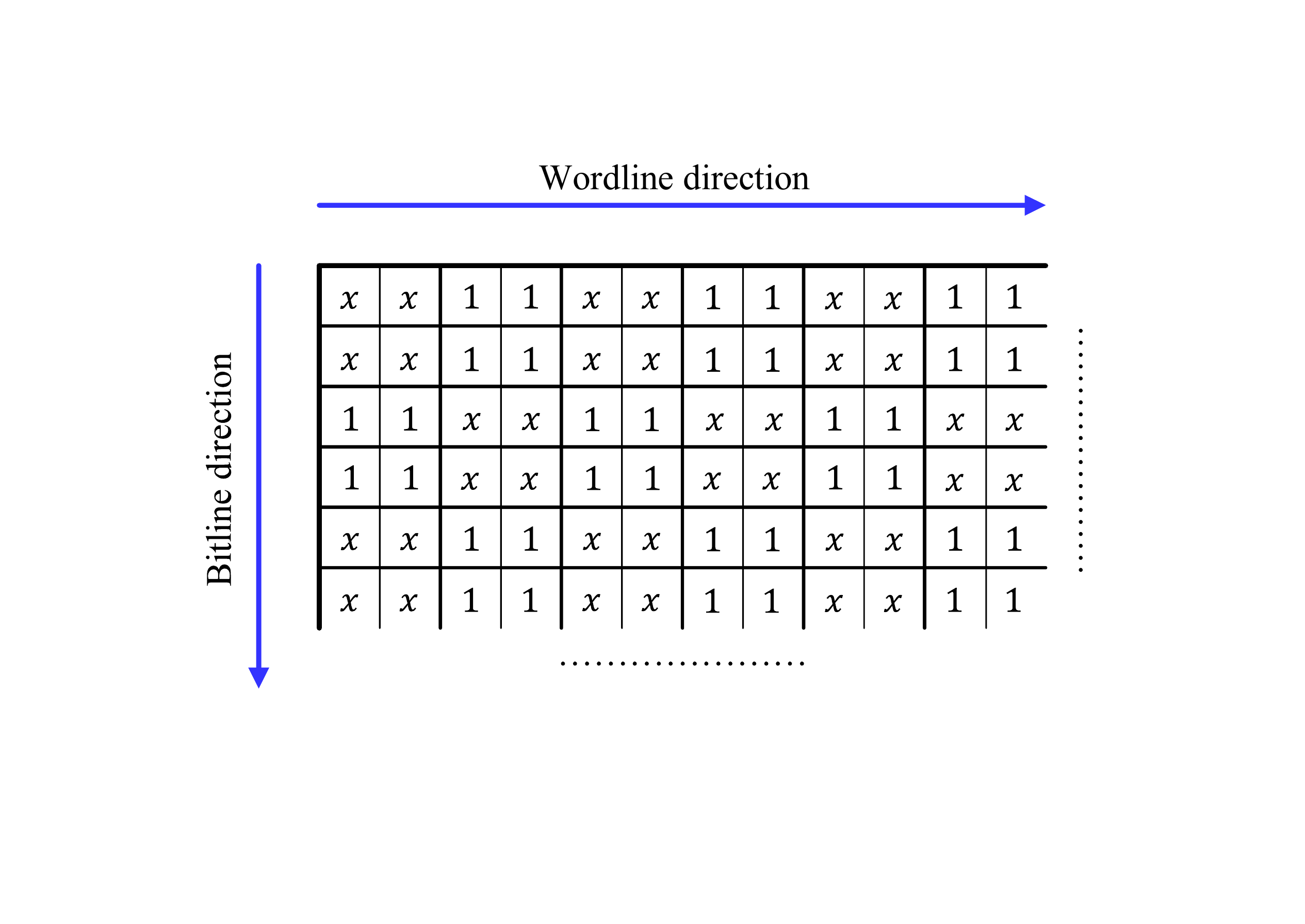}
\vspace{-0.5em}
\caption{The left-most pages of a 2D Flash grid with data encoded via the proposed 2D RR coding scheme. Symbol $x$ means bit can be $0$ or $1$ freely.}
\label{fig_1}
\vspace{-0.5em}
\end{figure}

\section{RR-LOCO Coding Over GF($2$)}\label{sec_rrloco2}

In this section, we introduce an RR coding scheme that forbids $\{000,010\}$ on the left-most pages in either the wordline direction or the bitline direction, while leaving all other pages with no coding, which forbids the level patterns in $\mathcal{L}_q$ and achieves page separation. The code we apply is a binary LOCO code devised according to the general method in \cite{ahh_general}. We start by defining the LOCO code.

\begin{definition}
A LOCO code $\mathcal{RC}_m$, where $m \geq 1$, that forbids $\{000,010\}$ is defined by the following properties:
\begin{enumerate}
\item Codewords in $\mathcal{RC}_m$ are defined over GF$(2) = \{0,1\}$ and are of length $m$ bits.
\item Codewords in $\mathcal{RC}_m$ are ordered lexicographically.
\item Codewords in $\mathcal{RC}_m$ do not have patterns in $\{000,010\}$.
\item All codewords satisfying 1)--3) are included.
\end{enumerate}
\end{definition}

Lexicographic ordering is ordering codewords ascendingly according to the rule ``$0 < 1$'', where bit significance reduces from left to right \cite{tang_bahl, ahh_qaloco}. The first step to devise the LOCO code is to specify the group structure. Codewords in $\mathcal{RC}_m$, $m \geq 2$, can be partitioned into the following groups:
\begin{itemize}
\item Group~1: Codewords starting with $0011$ from the left.
\item Group~2: Codewords starting with $011$ from the left.
\item Group~3: Codewords starting with $1$ from the left.
\end{itemize}

The second step is to enumerate the codewords, which is done by Theorem~\ref{thm_enum}. Let $N(m) \triangleq \vert \mathcal{RC}_m \vert$.

\begin{theorem}\label{thm_enum}
The cardinality of a LOCO code $\mathcal{RC}_m$ is given by the recursive formula:
\begin{equation}\label{eqn_enum}
N(m) = N(m-1)+N(m-3)+N(m-4), \text{ } m \geq 2,
\end{equation}
where the defined cardinalities and $N(1)$ are:
\begin{equation}\label{eqn_def}
N(-2) = N(-1) = N(0) \triangleq 1 \text{ and } N(1)=2.
\end{equation}
\end{theorem}

\begin{IEEEproof}
We compute the cardinalities of each group then add them all. Let the cardinality of Group~$i$ be $N_i$. As for Group~3 in $\mathcal{RC}_m$, there is a bijection between its codewords and the codewords in $\mathcal{RC}_{m-1}$ (attach $1$). Thus,
\begin{equation}\label{eqn_enum3}
N_3(m) = N(m-1).
\end{equation}
As for Group~2 in $\mathcal{RC}_m$, there is a bijection between its codewords and the codewords starting with $1$ from the left in $\mathcal{RC}_{m-2}$ (attach $01$). Thus using (\ref{eqn_enum3}),
\begin{equation}\label{eqn_enum2}
N_2(m) = N_3(m-2) = N(m-3).
\end{equation}
As for Group~1 in $\mathcal{RC}_m$, there is a bijection between its codewords and the codewords starting with $1$ from the left in $\mathcal{RC}_{m-3}$ (attach $001$). Thus using (\ref{eqn_enum3}),
\begin{equation}\label{eqn_enum1}
N_1(m) = N_3(m-3) = N(m-4).
\end{equation}
Adding (\ref{eqn_enum3}), (\ref{eqn_enum2}), and (\ref{eqn_enum1}) gives (\ref{eqn_enum}). The defined cardinalities can be computed by observing that $N(1)=2$, $N(2)=4$, and $N(3)=6$, which sets up three equations. This observation is immediate given the forbidden patterns.
\end{IEEEproof}

Define a codeword $\bold{c}$ in $\mathcal{RC}_m$ as $\bold{c} \triangleq c_{m-1} c_{m-2} \dots c_0$, with $c_i \triangleq \zeta$ for $i \geq m$, where $\zeta$ represents out of codeword bounds. The integer equivalent of a LOCO codeword bit $c_i$, $0 \leq i \leq m-1$, is $a_i$, i.e., $a_i$ is $0$ ($1$) when $c_i$ is $0$ ($1$). Denote the lexicographic index of a codeword $\bold{c}$ among all codewords in the LOCO code $\mathcal{RC}_m$ by $g(\bold{c})$. In general, $g(\bold{c})$ is in $\{0, 1, \dots, N(m)-1\}$.

The third step is to specify the special cases of occurence for a $1$ inside a codeword in $\mathcal{RC}_m$. These cases are:
\begin{itemize}
\item Case 1: $c_{i+2}c_{i+1}c_i = 001$.
\item Case 2: $c_{i+2}c_{i+1}c_i = 011$.
\item Case 3: $c_{i+2}c_{i+1}c_i = 101$ or $c_{i+2}c_{i+1}c_i = \zeta 01$.
\end{itemize}
The typical or default case is simply the case of ``otherwise''. In particular, it is the case that $c_{i+2}c_{i+1}c_i=111$, $c_{i+2}c_{i+1}c_i=\zeta 11$, or $c_{i+1}c_i=\zeta 1$.

The fourth and fifth steps are to find the encoding-decoding rule, which specifies the mapping from index to codeword and vice versa. This rule for $\mathcal{RC}_m$ is given in Theorem~\ref{thm_rule}.

\begin{theorem}\label{thm_rule}
The relation between the lexicographic index $g(\bold{c})$, $\bold{c} \in \mathcal{RC}_m$, and the codeword $\bold{c}$ itself is given by:
\begin{align}\label{eqn_rule}
g(\bold{c}) = \sum_{i=0}^{m-1} a_i &\Big [ (1-y_{i,1})N(i-2) \nonumber \\ &+ (1-y_{i,1}-y_{i,2})N(i-3) \Big ],
\end{align}
where $y_{i,1}$ and $y_{i,2}$ are specified as follows:
\begin{align}\label{eqn_rdef}
y_{i,1} &= 1 \text{ if } c_{i+2} c_{i+1} c_i \in \{001,011\}, \text{ and } y_{i,1} = 0 \text{ otherwise}, \nonumber \\
y_{i,2} &= 1 \text{ if } c_{i+2} c_{i+1} c_i \in \{101,\zeta01\}, \text{ and } y_{i,2} = 0 \text{ otherwise}.
\end{align}
\end{theorem}

\begin{IEEEproof}
We compute the contributions $g_{i,j}(c_i)$ of a bit $c_i$ under Case $j$, for all $j \in \{0,1,2,3\}$, in a LOCO codeword then merge them all. As for the typical case, which we index by $0$, this contribution is the number of codewords starting with $0$ from the left in $\mathcal{RC}_{i+1}$. Thus using (\ref{eqn_enum2}) and (\ref{eqn_enum1}),
\begin{align}\label{eqn_rule0}
g_{i,0}(c_i) &= N_2(i+1) + N_1(i+1) \nonumber \\ &= N(i-2) + N(i-3).
\end{align}
As for Case 1 (Case 2), this contribution is the number of codewords starting with $000$ ($010$) from the left in $\mathcal{RC}_{i+3}$. Note that $000$ and $010$ are forbidden patterns. Thus,
\begin{align}\label{eqn_rule12}
g_{i,1}(c_i) &= 0 \textup{ and } \nonumber \\ g_{i,2}(c_i) &= 0.
\end{align}
As for Case 3, this contribution is the number of codewords starting with $00$ from the left in $\mathcal{RC}_{i+2}$. Thus using (\ref{eqn_enum1}),
\begin{equation}\label{eqn_rule3}
g_{i,3}(c_i) = N_1(i+2) = N(i-2).
\end{equation}
Using $y_{i,1}$ and $y_{i,2}$ from (\ref{eqn_rdef}) along with $a_i$ to merge (\ref{eqn_rule0}), (\ref{eqn_rule12}), and (\ref{eqn_rule3}) gives:
\begin{equation}\label{eqn_girule}
g_i(c_i) = a_i \Big [ (1-y_{i,1})N(i-2) + (1-y_{i,1}-y_{i,2})N(i-3) \Big ].
\end{equation}
Substituting (\ref{eqn_girule}) in $g(\bold{c}) = \sum_{i=0}^{m-1} g_i(c_i)$ gives (\ref{eqn_rule}).
\end{IEEEproof}

For brevity, we skip the sixth step, which is to assemble the encoding and decoding algorithms. These algorithms are a direct consequence of the rule in (\ref{eqn_rule}), and we refer the reader to \cite{tang_bahl}, \cite{ahh_qaloco}, \cite{ahh_general}, and \cite{laroia_const} for details. Note that we sometimes refer to $\mathcal{RC}_m$ as a \textit{1D RR-LOCO code}. The encoding-decoding rule of a LOCO code is the reason behind its low complexity algorithms, where reconfiguration becomes as easy as reprogramming an adder \cite{ahh_loco, ahh_general}.

\begin{remark}
If the coded bits are complemented before writing to pages, the set of forbidden patterns on the left-most pages becomes $\{101,111\}$ instead, which appears in \cite{veeresh_mlc} as well. In this case, the cardinality of the LOCO code remains as in (\ref{eqn_enum}), while the encoding-decoding rule becomes exactly that of an asymmetric LOCO code in \cite{ahh_aloco} for $x=1$:
\vspace{-0.3em}\begin{equation}
g(\bold{c}) = \sum_{i=0}^{m-1} a_i N(i-a_{i+1}).\vspace{-0.1em}
\end{equation}
\end{remark}

Encoding and decoding on the left-most pages are just subtractions and additions. As for the remaining pages, data is written and read directly. This guarantees simplicity and maintains high access speed via our 1D RR-LOCO coding~scheme.


\section{Rate, Complexity, and Error Propagation}\label{sec_compropag}

We start by calculating asymptotic rates. Unfortunately, deriving the capacity for 2D constrained codes is known to be notoriously hard. Therefore, we will derive the capacity $C^{\textup{1D}}_{\mathcal{L}_q}$ only under the 1D constrained coding setup, which is already higher than the capacity under the 2D setup. Thus, $C^{\textup{1D}}_{\mathcal{L}_q}$ serves as a ceiling for the highest achievable rate in a device where patterns in $\mathcal{L}_q$ are forbidden at least in one direction. We will shortly show that 1D constrained coding suffices.

An FSTD of a sequence where level patterns in $\mathcal{L}_q$ are forbidden is shown in Fig.~\ref{fig_2}. Based on this FSTD, the general adjacency matrix is:
\vspace{-0.1em}\begin{equation}\label{eqn_adjmax}
\bold{A}_1=\left[
\begin{array}{c|c|c|c}
\frac{q}{2} & \bold{1}^\mathrm{T}_{\frac{q}{2}} & 0 & \bold{0}^\mathrm{T}_{\frac{q}{2}-1} \\ \hline \vspace{-0.9em} &  &  & \\
\bold{0}_{\frac{q}{2}} & \bold{U}^1_{\frac{q}{2}} & \frac{q}{2} \bold{1}_{\frac{q}{2}} & \makecell{\bold{0}^\mathrm{T}_{\frac{q}{2}-1} \\ \hline \bold{L}^1_{\frac{q}{2}-1}} \\ &  &  & \vspace{-0.9em} \\ \hline
\frac{q}{2} & \bold{0}^\mathrm{T}_{\frac{q}{2}} & 0 & \bold{0}^\mathrm{T}_{\frac{q}{2}-1} \\ \hline \vspace{-0.9em} &  &  & \\ 
\bold{0}_{\frac{q}{2}-1} & \begin{array}{c|c} \hspace{-0.5em} \bold{I}_{\frac{q}{2}-1} & \bold{0}_{\frac{q}{2}-1 \hspace{-0.5em}} \end{array} & \frac{q}{2} \bold{1}_{\frac{q}{2}-1} & \makecell{\bold{0}^\mathrm{T}_{\frac{q}{2}-1} \\ \hline \begin{array}{c|c} \hspace{-0.5em} \bold{L}^1_{\frac{q}{2}-2} & \bold{0}_{\frac{q}{2}-2 \hspace{-0.5em}} \end{array}} \vspace{-0.9em} \\ &  &  &
\end{array}\right].
\end{equation}
Thus and from \cite{shan_const}, the normalized capacity of a 1D constrained code forbidding the level patterns in $\mathcal{L}_q$ is:
\begin{equation}
C^{\textup{1D}}_{\mathcal{L}_q} = \frac{\log_2 (\lambda_{\max}(\bold{A}_1))}{\log_2 q},
\end{equation}
where $\lambda_{\max}(\bold{A})$ is the maximum real positive eigenvalue of the matrix~$\bold{A}$.\footnote{For positive integers $a+b \leq q$, the set $H$ of the $a$ largest symbols and the set $L$ of the $b$ smallest symbols in $\{0, 1, \ldots, q-1\}$, a formula for the (count-constrained) capacity of the constrained system forbidding patterns $\{\beta_1 \beta_2 \beta_3 \text{ } \vert \text{ } \beta_1,\beta_3\in H, \beta_2\in L\}$ was derived in~\cite{kashyap_isit2019}.}

\begin{figure}
\vspace{-0.5em}
\center
\includegraphics[trim={0.8in 1.4in 1.4in 1.0in}, width=3.5in]{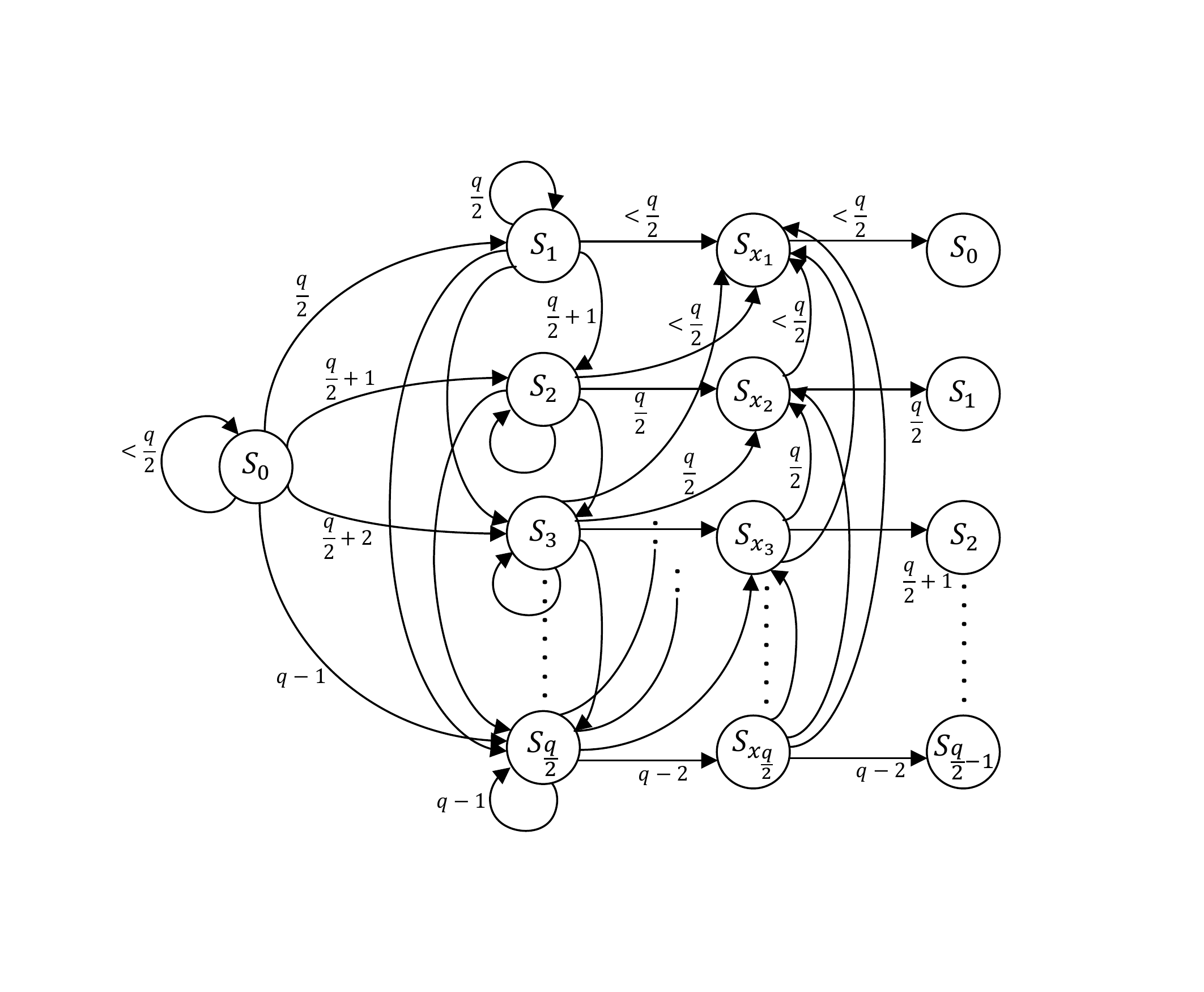}
\vspace{-0.5em}
\caption{An FSTD of a constrained sequence forbidding level patterns in $\mathcal{L}_q$, for any $q$. We operate directly on level patterns for simplicity.}
\label{fig_2}
\vspace{-0.3em}
\end{figure}

\begin{table}
\caption{Capacity Gap Between $C^{\textup{1D}}_{\mathcal{L}_q}$ and 1D RR-LOCO Capacity $C^{\textup{1D}}_{\textup{RR}}$}
\vspace{-0.5em}
\centering
\scalebox{1.00}
{
\begin{tabular}{|c|c|c|c|}
\hline
$q$ & \makecell{$C^{\textup{1D}}_{\mathcal{L}_q}$} & \makecell{$C^{\textup{1D}}_{\textup{RR}}$} & \makecell{Capacity gap $\%$} \\
\hline
$4$ & $0.8941$ & $0.8471$ & $5.257\%$ \\
\hline
$8$ & $0.9235$ & $0.8981$ & $2.750\%$ \\
\hline
$16$ & $0.9401$ & $0.9235$ & $1.766\%$ \\
\hline
\end{tabular}}
\label{table_1}
\vspace{-0.5em}
\end{table}

The capacity of a 2D code preventing $\{000,010\}$ is the capacity of a 2D $(0,1)$ RLL code, which is $\approx 0.5879$ \cite{kato_tcon}. Thus, the normalized capacity of our 2D RR coding scheme~is:
\begin{equation}
C^{\textup{2D}}_{\textup{RR}} \approx \frac{0.5879+\log_2 q -1}{\log_2 q} = \frac{\log_2 q - 0.4121}{\log_2 q}.
\end{equation}

As mentioned above, the constrained system 
where patterns in $\{000,010\}$ are forbidden 
can be interpreted as an interleaved RLL $(d,k)=(0,1)$ constraint, whose capacity is known to be $\log_2((1+\sqrt{5})/2)\approx 0.6942$. Thus, the normalized capacity of our 1D RR-LOCO coding scheme is 
\vspace{-0.1em}\begin{equation}
C^{\textup{1D}}_{\textup{RR}} = \frac{\log_2((1+\sqrt{5})/2)+\log_2 q -1}{\log_2 q} \approx \frac{\log_2 q - 0.3058}{\log_2 q}.
\end{equation}

The capacity gap between $C^{\textup{1D}}_{\mathcal{L}_q}$ and $C^{\textup{1D}}_{\textup{RR}}$ for different values of $q$ is given in Table~\ref{table_1}. The table shows that the capacity gap is small, and it gets even smaller as $q$ increases.

Next, we discuss the finite-length rates. First, the normalized rate of our 2D RR constrained coding scheme is:
\begin{equation}\label{eqn_rate2d}
R^{\textup{2D}}_{\textup{RR}} = \frac{0.5+\log_2 q -1}{\log_2 q} = \frac{\log_2 - 0.5}{\log_2 q}
\end{equation}
since the rate of our left-most page coding is $0.5$.

Regarding our 1D RR-LOCO coding scheme, we bridge with the pattern $11$ between consecutive codewords in $\mathcal{RC}_m$ on the left-most page, and we remove the codeword $\bold{1}^m$ for self-clocking \cite{ahh_qaloco, ahh_general}. Thus, the rate on the left-most page is $\lfloor \log_2 (N(m)-1) \rfloor/(m+2)$, and the normalized rate of our 1D RR-LOCO coding scheme is:
\begin{equation}\label{eqn_rate1d}
R^{\textup{1D}}_{\textup{RR}} = \frac{1}{\log_2 q} \left [ \frac{\lfloor \log_2 (N(m)-1) \rfloor}{m+2}+\log_2 q -1 \right ].
\end{equation}

1D RR-LOCO coding schemes are capacity-achieving schemes in the sense that the limit as $m \rightarrow \infty$ of $R^{\textup{1D}}_{\textup{RR}}$ is $C^{\textup{1D}}_{\textup{RR}}$ (see also \cite{ahh_qaloco}). Another capacity-achieving 1D RR constrained coding scheme, implementable using enumerative coding without the need for bridging bits, can be obtained by interleaving codewords from an optimal block code for the RLL $(d,k)=(0,1)$ constraint~\cite{marcus_jsac1992} on the left-most pages. LOCO codes, however, offer simplicity and reconfigurability, which is important as the device ages \cite{ahh_qaloco}.

The 2D RR constrained coding scheme we propose requires no additional complexity for encoding and decoding since data is written/read directly to/from pages. As for the 1D RR-LOCO coding scheme, the complexity is governed by the size of the adder that executes the encoding-decoding rule, which is:
\begin{equation}\label{eqn_msg}
s = \lfloor \log_2 (N(m)-1) \rfloor
\end{equation}
bits. For ease of implementation and to avoid affecting the access speed, we prefer to apply the 1D RR-LOCO coding scheme along wordlines instead of bitlines since the performance is almost the same, as demonstrated by the experimental results in Section~\ref{sec_resultstlc}.

Similarly, the 2D RR coding scheme does not incur any error propagation. Thus, the error propagation factor of it is $E^{\textup{2D}}_{\textup{RR}}=1$. As for the 1D RR-LOCO coding scheme, there is no codeword-to-codeword error propagation. However, there exists limited error propagation resulting from the codeword-to-message conversion \cite{ahh_loco, ahh_qaloco} on the left-most page only. This error propagation reaches $\frac{s}{2}$ bits on average, where $s$ is the message length as well from (\ref{eqn_msg}). Consequently, the error propagation factor averaged over $\log_2 q$ pages is:
\begin{equation}
E^{\textup{1D}}_{\textup{RR}} = \frac{1}{\log_2 q} \left [ \frac{s}{2}+\log_2 q-1 \right ].
\end{equation}

\begin{table}
\caption{Rate, Complexity, and Error Propagation Comparisons Between 2D and 1D RR Constrained Coding}
\vspace{-0.5em}
\centering
\scalebox{1.00}
{
\begin{tabular}{|c|c|c|c|c|c|c|}
\hline
$q$ & $m$ & $R^{\textup{2D}}_{\textup{RR}}$ & $R^{\textup{1D}}_{\textup{RR}}$ & $s$ & $E^{\textup{2D}}_{\textup{RR}}$ & $E^{\textup{1D}}_{\textup{RR}}$ \\
\hline
$4$ & $7$ & $0.7500$ & $0.7778$ & $5$ & $1.000$ & $1.750$ \\
\hline
$4$ & $11$ & $0.7500$ & $0.8077$ & $8$ & $1.000$ & $2.500$ \\
\hline
$4$ & $21$ & $0.7500$ & $0.8261$ & $15$ & $1.000$ & $4.250$ \\
\hline
$8$ & $7$ & $0.8333$ & $0.8519$ & $5$ & $1.000$ & $1.500$ \\
\hline
$8$ & $11$ & $0.8333$ & $0.8718$ & $8$ & $1.000$ & $2.000$ \\
\hline
$8$ & $21$ & $0.8333$ & $0.8841$ & $15$ & $1.000$ & $3.167$ \\
\hline
$16$ & $7$ & $0.8750$ & $0.8889$ & $5$ & $1.000$ & $1.375$ \\
\hline
$16$ & $11$ & $0.8750$ & $0.9038$ & $8$ & $1.000$ & $1.750$\\
\hline
$16$ & $21$ & $0.8750$ & $0.9130$ & $15$ & $1.000$ & $2.625$ \\
\hline
\end{tabular}}
\label{table_2}
\vspace{-0.5em}
\end{table}

Table~\ref{table_2} gives the rates, adder sizes, and error propagation factors of the proposed RR schemes under various parameters. The 1D RR-LOCO coding scheme has a remarkable rate~advantage that reaches $10.147\%$, $6.096\%$, and $4.343\%$ for $q=4$, $q=8$, and $q=16$, respectively over the 2D RR constrained coding scheme. The 2D RR scheme has a clear advantage in terms of both complexity and error propagation as it requires no processing to encode and decode. Having said that, the error propagation factor of the 1D RR scheme decreases notably as $q$ increases. For example, $E^{\textup{1D}}_{\textup{RR}}=2.625$ for $q=16$ and $m=21$, which is remarkably small given the code length.

The two coding schemes can be used in the same device, but at different lifetime stages. The 1D RR-LOCO coding scheme can be used when the device is still fresh or until a moderate number of program/erase (P/E) cycles, while the 2D RR constrained coding scheme can be used when the~device ages, where preventing the error-prone patterns in both directions could make a difference and the rate loss could be acceptable. However, this performance difference is shown to be small in Section~\ref{sec_resultstlc}, at least for the TLC Flash device we used.

\begin{remark}
An idea that allows page separation for MLC Flash was introduced in \cite{veeresh_mlc}. However, the rate offered is only $0.7500$, which is significantly below the rates offered via our 1D RR coding scheme for MLC. Another idea that allows page separation for TLC Flash was introduced in \cite{ravi_const}. However, it only heuristically addresses the level pattern $707$. 
\end{remark}

\section{Experimental Results on TLC Flash}\label{sec_resultstlc}

To characterize the performance of the RR constrained coding schemes, we conducted program/erase (P/E) cycling experiments on several blocks of a commercial 1X-nm TLC Flash chip, as follows:
\begin{enumerate}
\item Erase Flash memory block under test.
\item Program all pages of block under test with data. For uncoded experiments, program pseudo-random data at each P/E cycle. For RR experiments, program prepared data satisfying RR constraints at each P/E cycle. 
\item For each successive P/E cycle of RR experiments, ``rotate'' the data, so the data that was written on the page $i$ is written on the page $(i+1)$, wrapping around the last page to the first page. 
\item Record bit errors and compute channel bit error rate (BER) every $100$ P/E cycles.
\end{enumerate}

\begin{remark}
Gray mappings used in Flash devices may vary between manufacturers and product generations.  The forbidden patterns can usually be modified in accordance with the mapping so that RR coding on one page per wordline will eliminate all or most of the patterns in $\mathcal{L}_q$ that induce the most severe ICI.
\end{remark}


Fig.~\ref{exp_1DRR} presents the channel BER from P/E cycle $0$ to P/E cycle $10{,}000$ using pseudo-random data and a rate $24{:}36$ 1D RR-LOCO code either along wordlines or along bitlines. Using (\ref{eqn_rate1d}), $R^{\textup{1D}}_{\textup{RR}} = 0.8889$. Therefore, the 1D coding scheme achieves about $99\%$ ($96\%$) of the capacity $C^{\textup{1D}}_{\textup{RR}}$ ($C^{\textup{1D}}_{\mathcal{L}_q}$). The uncoded performance is better than that of the RR codes up to around $1{,}800$ P/E cycles  and is notably worse thereafter. 


At the later stages of P/E cycling, ICI increases, and the results in Fig. \ref{exp_1DRR} reflect this phenomenon. Specifically, RR-LOCO codes along wordlines increase device lifetime by about $1{,}200$ P/E cycles when channel BER is $2\times 10^{-3}$, i.e., $37\%$ lifetime gain, and achieve about $2{,}600$ P/E cycles gain when channel BER is $3\times 10^{-3}$, i.e., $58\%$ lifetime gain. After $2{,}000$ P/E cycles, the BER of 1D RR coding is almost the same on wordlines as it is on bitlines.

Fig.~\ref{exp_allcodes} compares the BER performance of the $24{:}36$ 1D RR-LOCO code with the interleaved $12{:}18$ RLL $(d,k)=(0,1)$ code (which has an overall block length $36$ after interleaving) and the 2D RR code. Using (\ref{eqn_rate2d}), $R^{\textup{2D}}_{\textup{RR}} = 0.8333$. Therefore, the~2D coding scheme achieves about $93\%$ ($90\%$) of the capacity $C^{\textup{2D}}_{\textup{RR}}$ ($C^{\textup{1D}}_{\mathcal{L}_q}$). The two 1D RR coding schemes have similar performance, with both showing that bitline coding performs almost the same as wordline coding. The 2D RR coding scheme achieves a slightly better performance than the two 1D RR coding schemes.

An examination of level probabilities induced by 1D RR constraints builds intuition towards the experimental results in Figs.~\ref{exp_1DRR} and~\ref{exp_allcodes}. The probabilities of binary symbols $0$ and $1$ under the RLL $(d,k)=(0,1)$ constraint are approximately $0.2764$ and $0.7236$, respectively \cite{siegel_nasit2015}. Asymptotically, this leads to probabilities of individual symbols in $\mathcal{V}_0$ and $\mathcal{V}_1$ of about $0.0691$ and $0.1809$, respectively.
 
The cross-over behavior observed in Fig.~\ref{exp_1DRR} can be explained if the level patterns eliminated by the code, especially ICI-prone patterns, are not the only significant contributors to error early in the device lifetime. The RR coding significantly changes level probabilities compared with the uncoded setting, increasing the probability of some of the remaining level patterns that cause errors due to other effects, and accordingly increasing their contribution to the BER at low P/E cycles. One suggestion to prevent this behavior is applying two different constraints before and after the cross-over point, making use of the LOCO reconfigurability feature that could be directed by a machine learning module.
 
Similarly, 1D RR coding in the wordline direction will reduce the probabilities of detrimental patterns in the bitline direction, and vice versa. This reduces the impact of the more severe ICI in the bitline direction on the overall error rate, while simultaneously reducing the expected advantage of the 2D RR coding (even without taking into account the rate penalty associated with the 2D coding). 
 
These effects on level-pattern probabilities have been confirmed by examination of the data written to the Flash memory block and the observed error-inducing patterns. 

\begin{figure}
\center
\includegraphics[trim={0.0in 1.1in 0.0in 1.2in}, width=3.5in]{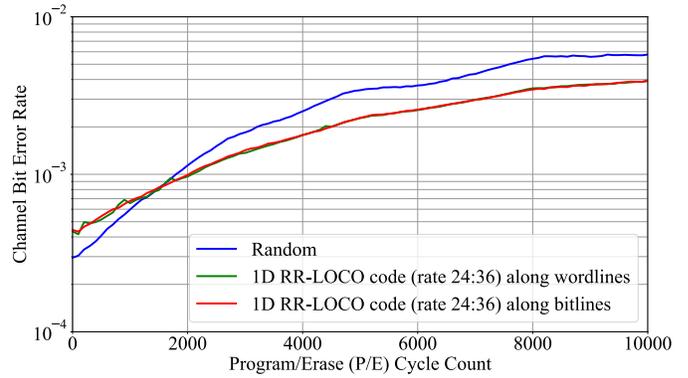}\vspace{-0.7em}
\caption{Measured average channel BER comparison when all pages are programmed with random data, 1D RR-LOCO coded data (rate $24{:}36$) along wordlines or bitlines.}
\vspace{-0.5em}
\label{exp_1DRR}
\end{figure}

\begin{figure}
\center
\includegraphics[trim={0.0in 1.25in 0.0in 1.0in}, width=3.5in]{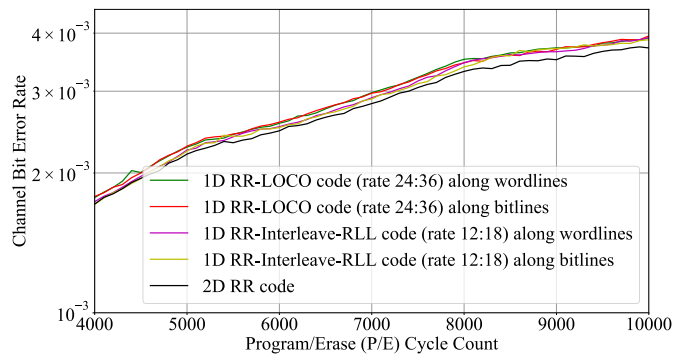}\vspace{-0.7em}
\caption{Measured average channel BER comparison of 1D RR-LOCO codes (rate $24{:}36$) along wordlines or bitlines, 1D interleaved RLL-$(0,1)$ codes (rate $12{:}18$) along wordlines or bitlines, and 2D RR code.}
\vspace{-0.5em}
\label{exp_allcodes}
\end{figure}

\section{Conclusion}\label{sec_concl}

We introduced 2D and 1D RR coding schemes for modern Flash devices. RR coding schemes are systematic, and they incur limited redundancy to improve performance. Experimental results reveal significant P/E cycles gains in a commercial Flash device. In summary, RR codes offer an efficient and practical approach to mitigating ICI that can enhance Flash device lifetime. Future work includes combining RR constrained codes with effective LDPC codes \cite{ahh_md}.




\end{document}